\documentclass[runningheads]{llncs}
\usepackage{makeidx}
\usepackage{multirow}
\usepackage{multicol}
\usepackage[dvipsnames,svgnames,table]{xcolor}
\usepackage{graphicx}
\usepackage{epstopdf}
\usepackage{ulem}
\usepackage{hyperref}
\usepackage{amsmath}
\usepackage{ragged2e}
\usepackage{amssymb}
\usepackage{graphicx}
\usepackage{enumerate}

\renewcommand{\raggedright}{\leftskip=0pt\rightskip=0pt}
\begin{document}
\title{A Hybrid Quantum Secret Sharing Scheme based on Mutually Unbiased Bases}
\titlerunning{QSS}
\makeatletter
	\newenvironment{indentation}[3]%
	{\par\setlength{\parindent}{#3}
	\setlength{\leftmargin}{#1}       \setlength{\rightmargin}{#2}%
	\advance\linewidth -\leftmargin       \advance\linewidth -\rightmargin%
	\advance\@totalleftmargin\leftmargin  \@setpar{{\@@par}}%
	\parshape 1\@totalleftmargin \linewidth\ignorespaces}{\par}%
\makeatother
\author{Dan-Li Zhi \and Zhi-Hui Li\thanks{lizhihui@snnu.edu.cn} \and Li-Juan Liu \and Zhao-Wei Han}
\authorrunning{D.L. Zhi et.al}
%
\institute{Shaanxi Normal University, xi'an, Shaanxi, 710119, China}

\maketitle              
\begin{abstract}
 With the advantages of both classical and quantum secret sharing, many practical hybrid quantum secret sharing have been proposed. In this paper, we propose a hybrid quantum secret sharing scheme based on mutually unbiased bases and monotone span program. First, the dealer sends the shares in the linear secret sharing to the participants in the authorization set via a secure channel. Then, the dealer and participants perform unitary transformation on a $d$-dimensional quantum state sequentially, and  the dealer publishes the measurement result confidentially to the participants in the authorization set to recover the secret. The verifiability of the scheme is guaranteed by the Hash function. Next, the correctness and security of the scheme are proved and our scheme is secure against the general eavesdropper attacks. Finally, a specific example is employed to further clarify the flexibility of the scheme and the detailed comparison of similar quantum secret sharing schemes also shows the superiority of our proposed scheme.
\end{abstract}
\section{Introduction}
As a combination of cryptography and quantum mechanics, quantum
cryptography plays an important role in cryptography. Compared with classical
cryptography on the basis of computational complexity, quantum cryptography based on the
laws of quantum physics can achieve unconditional security. Many branches of
quantum cryptography have been developed, such as quantum key
distribution(QKD)[1,2], quantum key agreement(QKA)[3-5], quantum secure direct
communication(QSDC)[6,7], quantum teleportation[8,9], quantum signature[10,11],
quantum authentication[12-14], quantum secret sharing(QSS)[15-30] and so on.

Quantum secret sharing (QSS) is an important research field in
quantum cryptography, which means that the dealer divides a secret into several
shadows and sends them to multiple participants. Only the participants in
authorized sets can recover the secret, and the participants in unauthorized sets
can not recover the secret. Since Hillery et al. [15] proposed the first quantum
secret sharing scheme by using GHZ state in 1999, a growing number of QSS schemes
[16-30] have been proposed. For example, Williams et al. [22] described and
experimentally demonstrated a three-party quantum secret sharing protocol using
polarization-entangled photon pairs. Tsai et al. [23] used the entanglement
property of W-state to propose the first three-party
SQSS protocol. Song et al. [24] demonstrated a $\left( {t,n} \right)$ threshold
$d$-level quantum secret sharing scheme. A verifiable $\left( {t,n} \right)$
threshold quantum secret sharing scheme was proposed using the $d$-dimensional
Bell state and the Lagrange interpolation by Yang et al. in Ref. [25]. Hao et al. [26]
put forward a secret sharing scheme using the mutually unbiased bases on
the $p^2$-dimensional quantum system.
Bai et al. [27] proposed the concept of decomposition of quantum access structure
to design a quantum secret sharing scheme.
In Ref. [28], Liu et al. study the local distinguishability of the 15 kinds of seven-qudit
quantum entangled states and then proposed a $(k,n)$ threshold quantum secret sharing scheme.
A new improving quantum secret sharing scheme was proposed by Xu et al. [29], in which more
quantum access structures can be realized by the scheme than the one proposed by Nascimento et al. [30].

Although many schemes have been proposed, the verifiability
and the flexibility of the schemes are also important issues
worth  of consideration. In this paper, we propose a hybrid and verifiable quantum secret sharing
scheme based on mutually unbiased bases and the monotone span program, which focuses on transmitting a
$d$-dimensional quantum state among the dealer Alice and participants and the application of the linear secret sharing. Each participant
in a authorization set can perform a unitary transformation on the received particle
and send it to the next one until the last one sends it to Alice.
They can recover the secret  by the linear secret sharing and the measurement value sent by Alice. Verifiability ensures that the secret recovered in each authorization set is the
original one, and it also ensures that once a dishonest participant appears, he
will be found. Compared with the threshold scheme, the quantum secret sharing
scheme based on the access structure realizes the different influences of
participants in the process of recovering secrets, thereby achieving the
flexibility of the scheme.

 By comparison, our scheme shows all the advantages of the previous QSS and the unique advantages, such as, \begin{enumerate}[(1)]
    \item {It uses a qudit state instead of a qubit state.}
    \item {The participants can check the authenticity of the recovered secret.}
    \item {It needs fewer quantum resources and quantum operations.}
    \item {It has general access structure.}
    \item {It reduces the communication costs and  computation complexity.}
\end{enumerate}

This paper is organized as follows. In section 2, we illustrate the preliminary
knowledge related to the proposed scheme. The new proposed scheme is introduced
in section 3. Section 4 give a proof of the correctness, verifiability and
security of the proposed scheme. In section 5, we give an example to further
illustrate our proposed scheme. Finally, the
comparison and conclusion is given in section 6 and section 7.

\section{Preliminaries}
In this section, we introduce the preliminary knowledge of our
scheme.
\subsection{Access structure}

\textbf{Defination 1} Let ${\cal P} = \left\{ {P_1,P_2, \cdots ,P_n}\right\}$
be a set of participants, an access structure $\Gamma  \subseteq {2^{\cal
P}}$ is a family of authorized sets of participants.

{\noindent{\textbf{Defination 2} If $\Gamma $ is the access structure on ${\cal P}$, then any
set in $\Gamma $ is called the authorization subset on ${\cal P}$, which is called
the authorization set for short. If $A \in \Gamma ,A \subseteq B \subseteq {\cal P}$,
then $B \in \Gamma ${\scriptsize .} The family of the unauthorized sets is called an
adversary structure, that is to say, ${\Gamma ^c}{\rm{ = }}\Delta
${\scriptsize .}}}

{\noindent{\textbf{Example 1} Let ${\cal P} = \left\{ {P_1},{P_2},{P_3},{P_4}\right\},
\Gamma  = \left\{ {{A_1},{A_2},{A_3}} \right\}$,
where ${A_1} = \left\{ {P_1},{P_2},{P_3} \right\}$,
${A_2} =\left\{ {P_1},{P_2},{P_4} \right\}$,
${A_3} = \left\{{P_1},{P_2},{P_3},{P_4}\right\}$. So
\[
{\Delta} = \left\{ \begin{array}{l}
\emptyset ,\left\{ {P_1} \right\},\left\{ {P_2} \right\},\left\{
{P_3} \right\},\left\{ {P_4} \right\},\left\{ {P_1,P_2}
\right\},\left\{ {P_1,P_3} \right\},\left\{ P_1,P_4\right\},\\
\left\{ {P_2,P_3} \right\},
\left\{ {P_2,P_4} \right\},\left\{ {P_3,P_4} \right\},\left\{{P_1,P_3,P_4} \right\}\left\{ {P_2,P_3,P_4} \right\}
\end{array} \right\}.
\]
}}
\subsection{Monotone span program}

MSP was introduced in Ref.[31] by Karchmer and Wigderson as a model of
computation to design the linear secret sharing scheme.

{\noindent{\textbf{Defination 3} ${\cal M}\left( {{\cal F},M,\psi ,\vec \xi } \right)$ is a monotone
span program(MSP), where $M$ is a $k \times l$ matrix over a finite
field ${\cal F}$, $\psi :\left\{ {1,2, \cdots ,k} \right\} \to {\cal P}$ is a
surjective labeling map, $\vec \xi  = {\left( {1,0, \cdots ,0}
\right)^T} \in {{\cal F}^l}$ is defined as the target vector. For any $A \subseteq {\cal P}
= $ $ \left\{ {P_1,P_2, \cdots ,P_n} \right\}$, there is a corresponding
eigenvector ${\vec \delta _A} = \left( {{{ \delta }_1},{{ \delta }_2},
\cdots ,{{ \delta }_n}} \right) \in {\left\{ {0,1} \right\}^n}$ if and only
if $P_i \in A,{\delta _i} = 1$. The Boolean function $f:{\left\{
{{\rm{0}},{\rm{1}}} \right\}^n} \to \left\{ {{\rm{0}},{\rm{1}}} \right\},f\left(
{{\delta _A}} \right) = 1$ represents the corresponding $\varepsilon ${\scriptsize
 } rows of $M$, where $\psi \left( \varepsilon  \right) \in A,\varepsilon  \in
\left\{ {1,2, \cdots ,k} \right\}${\scriptsize .}
}}

{\noindent{\textbf{Defination 4} A monotone span program (MSP) is called a MSP for access
structure $\Gamma $, if it can be satisfied that $\forall A \in \Gamma $,$\exists {\vec \lambda _A} \in {{\cal F}^k}$ $\Rightarrow$ $M_A^T{\vec
\lambda _A} = \vec \xi $, and $\forall A \in \Delta$ ,$\exists \vec h = \left( {1,{h_2}, \cdots ,{h_l}} \right)
\in {{\cal F}^l}$ $\Rightarrow$ $ M_A\vec h = \vec 0 \in {{\cal F}^m}$.
}}

{\noindent{\textbf{Example 2} ${\cal M}\left( {{\cal F},M,\psi ,\vec \xi } \right)$ is an MSP of
access structure $\Gamma $ as shown in example 1, where $ {\cal F}= {{\cal Z}_5}$,$\psi \left( i
\right) = Bo{b_i},i \in \left\{ {1,2,3,4} \right\}$,$\vec \xi  = {\left(
{1,0,0,0} \right)^T}$,$M = \left( {\begin{array}{*{20}{c}}
1&0&3&4\\
0&0&2&1\\
3&4&1&0\\
1&2&4&0
\end{array}} \right)$. Therefore, \textbf{{\scriptsize  ${\vec \lambda _{{A_1}}} =
\left( {1,1,0} \right)^T,{\vec \lambda _{{A_2}}} = \left( {1,1,0} \right)^T,{\vec \lambda
_{{A_3}}} = \left( {1,1,3,4} \right)^T$.}}
}}

\subsection{Linear secret sharing}

 Monotone span program is utilized to design the linear secret sharing scheme, which is aimed that the dealer Alice shares a secret
$s$ among $k$ shareholders $Bo{b_1},Bo{b_2}, \cdots ,Bo{b_k}$ according to the MSP for access structure $\Gamma
$. It includes the following two phases as follows.

{\noindent{\raggedright\textbf{Distribution phase}

{\raggedright Alice prepares a random vector $\vec \rho  = {\left( {s,{\rho _2}, \cdots ,{\rho
_l}} \right)^T} \in {^l}$ and computes $\vec s = M\vec \rho  = \left( {s{}_1,
\cdots ,s{}_k}\right)^T $. Then, she sends ${s_i}$ to $\psi \left( i \right)$ via a
secure channel.
}}}

{\noindent{\raggedright\textbf{Reconstruction phase}

{\raggedright Let ${\vec s_A}$ be indicated the vector for
the authorized set $A$. The participants in $A$ restore the secrets cooperatively as
follows.
}}}

\begin{equation}
\vec s_A^T{\vec \lambda _A} = {\left( {{M_A}\vec \rho } \right)^T}{\vec \lambda
_A} = {\vec \rho ^T}\left( {M_A^T{{\vec \lambda }_A}} \right) = {\vec \rho
^T}\vec \xi  = s.
\label{eq:1}
\end{equation}

\subsection{Necessary quantum properties}

{\raggedright\textbf{Defination 5} Mutually unbiased base is defined that two
sets of standard orthogonal bases ${A_1} = \left\{ {\left| {{\varphi _1}}
\right\rangle ,\left| {{\varphi _2}} \right\rangle , \cdots ,\left| {{\varphi
_d}} \right\rangle } \right\}$ and ${A_2} = \left\{ {\left| {{\psi _1}}
\right\rangle ,\left| {{\psi _2}} \right\rangle , \cdots ,\left| {{\psi _d}}
\right\rangle } \right\}$, which defined over a $d$-dimensional complex space
${C^d}$ in Ref.[32,33] , if the following relationship is satisfied

\begin{equation}
\left| {\left\langle {{\varphi _i}} \right.\left| {{\psi _i}} \right\rangle }
\right| = \frac{1}{{\sqrt d }}.
\label{eq:2}
\end{equation}

{\noindent{\raggedright
If any two of the set of standard orthogonal bases $\left\{ {{A_1},{A_2}, \cdots
,{A_m}} \right\}$ in space are unbiased, then this set is called an unbiased bases
set. Besides, it can be found $d + 1$\textbf{{\large  }} mutually unbiased bases
if $d$ is an odd prime number.
}}

{\noindent{\raggedright\textbf{Defination 6} The computation base is expressed as $\left\{ {\left| k
\right\rangle \left| {k \in D} \right.} \right\}$, and the remaining groups can be expressed as:

\begin{equation}
\left| {v_l^{\left( j \right)}} \right\rangle  = \frac{1}{{\sqrt d
}}\sum\limits_{k = 0}^{d - 1} {{w^{k\left( {l + jk} \right)}}\left| k
\right\rangle },
\label{eq:3}
\end{equation}

{\noindent where $\left| {v_l^{\left( j \right)}} \right\rangle$ represents the $l$-th vector
in the  $j$-th bases, $w = {e^{\frac{{2\pi i}}{d}}}l,j \in
D$, $D = \left\{ {0,1, \cdots ,d -
1} \right\}$. These mutually unbiased bases satisfy the
following conditions:
}

\begin{equation}
\left| {\left\langle {v_l^{\left( j \right)}} \right.\left| {v_l^{\left( {j'}
\right)}} \right\rangle } \right| = \frac{1}{{\sqrt d }},j \ne j'.
\label{eq:4}
\end{equation}
}}

{\noindent{\raggedright\textbf{Defination 7} In Ref.[34], the two unitary transformations ${X_d}$ and
${Y_d}$ that we need to use in this paper can be expressed as:

\begin{equation}
{X_d} = \sum\limits_{m = 0}^{d - 1} {{w^m}\left| m \right\rangle \left\langle m
\right|} ,{Y_d} = \sum\limits_{m = 0}^{d - 1} {{w^{{m^2}}}\left| m \right\rangle
\left\langle m \right|}.
\label{eq:5}
\end{equation}

{\noindent{\raggedright Implementing (5) on $\left| {v_l^{\left( j \right)}} \right\rangle $ in turn, we
can obtain:
}

\begin{equation}
X_d^xY_d^y\left| {v_l^{\left( j \right)}} \right\rangle  = \left| {v_{l +
x}^{\left( {j + y} \right)}} \right\rangle.
\label{eq:6}
\end{equation}

{\noindent For the convenience of expression, $X_d^xY_d^y$ is denoted as ${U_{x,y}},$ that
is,}

\begin{equation}
{U_{x,y}}\left| {v_l^{\left( j \right)}} \right\rangle  =
\left| {v_{l +x}^{\left( {j + y} \right)}} \right\rangle .
\label{eq:7}
\end{equation}
}}}
\section{Proposed scheme}

In this section, we construct a  verifiable quantum secret sharing scheme that includes a dealer Alice and $n$ shareholders $Bo{b_1},Bo{b_2}, \cdots ,Bo{b_n}$. The
access structure $\Gamma $ can be expressed as $\Gamma  = \left\{ {{A_1},{A_2},
\cdots ,{A_r}} \right\}$, where ${A_i}(i = 1,2, \cdots ,r)$  is a authorization set.
For the convenience of description, the authorization set is recorded as ${A_i} =
\left\{ {Bob_1^{(i)},Bob_2^{(i)}, \cdots ,Bob_m^{(i)}} \right\}$, $(1 \le m \le n)$.
Without losing generality, it is assumed that the participants in the authorization set ${A_i} =
\left\{ {Bob_1^{(i)},Bob_2^{(i)}, \cdots ,Bob_m^{(i)}} \right\}$ want to recover
the secret $s$. The specific steps of the scheme are as follows.

\subsection{Distribution phase}

Alice implements the following steps.

{\noindent{\raggedright\textbf{3.1.1} Select a random vector $\vec \rho  = {\left( {{S_i},{\rho _2},{\rho _3} \cdots ,{\rho
_l}} \right)^T}$ according to authorization set ${A_i}${\scriptsize .}}}

{\noindent{\raggedright\textbf{3.1.2} Calculate ${\vec s} = M_{n\times l}\vec \rho  = {\left( {s_1^{(i)},s_2^{(i)},,
\cdots ,s_n^{(i)}} \right)^T},i = 1,2, \cdots ,r$ and send $\vec s_j^{\left( i
\right)}$ to $\psi (j) = Bo{b_j}(j = 1,2, \cdots ,n)$ through the quantum secure channel.}}

{\noindent{\raggedright\textbf{3.1.3} Compute and publish $H_1 = h(S_i)$,$ H_2=h(s)$, where $h()$ is a public Hash function.}}

{\noindent{\raggedright\textbf{3.1.4} Prepare a quantum state $\left| \phi  \right\rangle  = \left| {\varphi _0^0}
\right\rangle  = \frac{1}{{\sqrt d }}\sum\limits_{j = 0}^{d - 1} {\left| j
\right\rangle } $ and perform a unitary operation ${U_{p_0^{(i)},q_0^{(i)}}}$to
get the quantum state $\left| \phi  \right\rangle _0^{(i)} =
{U_{p_0^{(i)},q_0^{(i)}}}\left| {\varphi _0^0} \right\rangle  = \left| {\varphi
_{p_0^{(i)}}^{q_0^{(i)}}} \right\rangle $, where $p_0^{(i)} = s$ is the secret, $q_0^{(i)}$ is a
secret value known only to Alice. Then, she sends the quantum state $\left|
\phi  \right\rangle _0^{(i)}$ performed by the unitary operation to the first
participant $Bob_1^{(i)}$ in the authorization set ${A_i}$.}}

\subsection{Reconstruction phase}

Participants in ${A_i} =\left\{ {Bob_1^{(i)},Bob_2^{(i)}, \cdots ,Bob_m^{(i)}} \right\},(1 \le m \le n)$ can recover the secret
 by the following steps.

{\noindent{\raggedright\textbf{3.2.1} After receiving the quantum state ${\left| \phi  \right\rangle _0}$, the first
participant $Bob_1^{(i)}$ performs unitary
operates ${U_{p_1^{(i)},q_1^{(i)}}}$on it and gets the quantum state $\left| \phi
\right\rangle _1^{(i)} = {U_{p_1^{(i)},q_1^{(i)}}}$ $\left| {\varphi
_{p_0^{(i)}}^{q_0^{(i)}}} \right\rangle  = \left| {\varphi _{p_0^{(i)} +
p_1^{(i)}}^{q_0^{(i)} + q_1^{(i)}}} \right\rangle $. Next, the quantum
state $\left| \phi  \right\rangle _1^{(i)}$ is sent to the second
participant $Bob_2^{(i)}$ in the authorization set ${A_i}$, where $p_1^{(i)} =
\lambda _1^{(i)}s_1^{(i)},q_1^{(i)} = \lambda _1^{(i)}${\scriptsize .  }}}

{\noindent{\raggedright\textbf{3.2.2} The other participants $Bob_j^{(i)}(j = 2,3, \cdots ,m)$ in the authorization set
${A_i}$ perform the same operation as in step 3.2.1, which means that after
receiving the quantum state $\left| \phi  \right\rangle _{j - 1}^{(i)}$,
$Bob_j^{(i)}$ performs unitary operation ${U_{p_j^{(i)},q_j^{(i)}}}$ on it and gets
the quantum state $\left| \phi  \right\rangle _j^{(i)} =
{U_{p_j^{(i)},q_j^{(i)}}}\left| {\varphi _{\sum\limits_{k = 0}^{j - 1}
{p_k^{(i)}} }^{\sum\limits_{k = 0}^{j - 1} {q_k^{(i)}} }} \right\rangle  = \left|
{\varphi _{\sum\limits_{k = 0}^j {p_k^{(i)}} }^{\sum\limits_{k = 0}^j {q_k^{(i)}}
}} \right\rangle $, and then sends it to the next participant $Bob_{j + 1}^{(i)},(j
= 2,3, \cdots m - 1)$ until the last participant $Bob_m^{(i)}$ in the
authorization set ${A_i}$ completes the operation and sends the final quantum
state to Alice, where $p_j^{(i)} = \lambda _j^{(i)}s_j^{(i)},q_j^{(i)} = \lambda _j^{(i)}${\scriptsize . }
For the authorization set ${A_i}$, when all the
participants act and transmit, the final quantum state is

\begin{equation}
\left| \phi  \right\rangle _m^{\left( i \right)} = \prod\limits_{k = 0}^m
{{U_{p_k^{(i)},q_k^{(i)}}}} \left| {\varphi _0^0} \right\rangle  = \left|
{\varphi _{\sum\limits_{k = 0}^m {p_k^{(i)}} }^{\sum\limits_{k = 0}^m {q_k^{(i)}}
}} \right\rangle.
\label{eq:8}
\end{equation}
}}

{\noindent{\raggedright\textbf{3.2.3} When Alice receives the final quantum state {\scriptsize  }$\left| \phi
\right\rangle _m^{\left( i \right)}$, she can know that it  satisfies the
following condition on account of $q_0^{(i)},q_1^{(i)}, \cdots ,q_m^{(i)}$,

\begin{equation}
q_0^{(i)} + q_1^{(i)} +  \cdots  + q_m^{(i)} = {q_i}.
\label{eq:9}
\end{equation}

{\noindent She selects the measurement bases ${M_{{q_i}}} = \left\{ {\left| {\varphi
_j^{({q_i})}} \right\rangle \left| {j \in D} \right.} \right\}$ to measure it,
and then infers the following condition should be established in the authorization set ${A_i}$

\begin{equation}
p_0^{(i)} + p_1^{(i)} +  \cdots  + p_m^{(i)} = p_0^{(i)}+S_i=r_i
\label{eq:10}
\end{equation}

{\noindent If it is established, Alice checks whether $H_1$ of the participants are equal to the published one. If so,
the measurement results ${r_i}$ will be sent to all participants in
the authorization set ${A_i}$ through  the secure channel and then it move to the next step. If not, the scheme is terminated.}}}

{\noindent{\raggedright\textbf{3.2.4} In order to reconstruct the secret, each participant in authorization set
${A_i}$ can recover the secret by calculating $s = {p_0} = {r_i} -
\sum\limits_{i = 1}^m {{p_i}}={r_i} -{S_i} $.}}

\section{Correctness, verifiability and security}
In this section, the provability of the correctness, verifiability and security of our scheme is given.
\subsection{Correctness}

\textbf{Theorem 1} If a $d$-dimensional quantum state in mutually unbiased bases
is $\left| {v_l^{\left( j \right)}} \right\rangle  = \frac{1}{{\sqrt d
}}\sum\limits_{k = 0}^{d - 1} {{w^{k\left( {l + jk} \right)}}\left| k
\right\rangle } $, and a unitary operation ${U_{x,y}} = X_d^xY_d^y$ is performed
on it, then it will become another state $\left| {v_{l + x}^{\left( {j + y}
\right)}} \right\rangle $, that is, ${U_{x,y}}\left| {v_l^{\left( j \right)}}
\right\rangle  = \left| {v_{l + x}^{\left( {j + y} \right)}} \right\rangle
$.
\textbf{Proof} When implementing $Y_d^y,X_d^x$ on $\left| {v_l^{\left( j
\right)}} \right\rangle $ in turn, we can obtain,

\begin{equation}
\begin{array}{l}
X_d^xY_d^y\left| {v_l^{\left( j \right)}} \right\rangle  = X_d^x\left(
{\sum\limits_{m = 0}^{d - 1} {{w^{y{m^2}}}\left| m \right\rangle \left\langle m
\right|} } \right)\left( {\frac{1}{{\sqrt d }}\sum\limits_{k = 0}^{d - 1}
{{w^{k\left( {l + jk} \right)}}\left| k \right\rangle } } \right)\\
\;\;\;\;\;\;\;\;\;\;\;\;\;\;\;\;\; \;\;\;= \frac{1}{{\sqrt d }}\sum\limits_{m = 0}^{d - 1}
{{w^{xm}}\left| m \right\rangle \left\langle m \right|} \sum\limits_{k = 0}^{d -
1} {{w^{k\left( {l + (j + y)k}\;\; \right)}}\left| k \right\rangle } \\
\;\;\;\;\;\;\;\;\;\;\;\;\;\;\; \;\;\;\;\;= \frac{1}{{\sqrt d }}\sum\limits_{k = 0}^{d - 1}
{{w^{k\left[ {(l + x) + (j + y)k} \right]}}\left| k \right\rangle } \\
\;\;\;\;\;\;\;\;\;\;\;\;\; \;\;\;\;\;\;\;= \left| {v_{l + x}^{\left( {j + y} \right)}}
\right\rangle .
\end{array}
\label{eq:11}
\end{equation} This completes the proof.

{\noindent\textbf{Lemma 1 }In the secret sharing scheme, according to Theorem 1, the initial
state selected by Alice is $\left| \phi  \right\rangle  = \left| {\varphi _0^0}
\right\rangle  = \frac{1}{{\sqrt d }}\sum\limits_{j = 0}^{d - 1} {\left| j
\right\rangle } $, and the unitary operation ${U_{p_k^{\left( i
\right)},q_k^{\left( i \right)}}} = X_d^{p_k^{\left( i \right)}}Y_d^{q_k^{\left(
i \right)}},k = 0,1, \cdots ,m${\scriptsize  } is performed on the states
sequentially by Alice and all the participants in the authorization set ${A_i}$, then the
final state is $\left| \phi  \right\rangle _m^{\left( i \right)} = \left(
{\prod\limits_{u = 0}^m {{U_{{p_u},{q_u}}}} } \right){\left| \phi  \right\rangle}$, that is, $\left| \phi  \right\rangle _m^{\left( i \right)} =
\prod\limits_{k = 0}^m {{U_{p_k^{(i)},q_k^{(i)}}}} \left| {\varphi _0^0}
\right\rangle  = \left| {\varphi _{\sum\limits_{k = 0}^m {p_k^{(i)}}
}^{\sum\limits_{k = 0}^m {q_k^{(i)}} }} \right\rangle $.When Alice announces the
measurement result ${r_i}$ via the quantum secure channel to the participants in ${A_i}$,
they can restore the secret $s = {p_0} = {r_i} -
\sum\limits_{k = 1}^m {{p_k^{(i)}}}={r_i} -{S_i} $.
}

\subsection{Verifiability}

On one hand, before Alice sends the measurement result, she can check $H_1$ to ensure that the secret value recovered by linear secret sharing is correct, which provides a prerequisite for participants to recover the correct secret.
On the other hand ,each participant can check

\begin{equation}
H_2= h\left( s \right),
\label{eq:12}
\end{equation}

{\noindent to ensure that the recovered secret is the original one.}
\subsection{Security}

We analyze the security of our scheme against the general attacks here.

\subsubsection{Entangle and measure attack}

{\raggedright We assume that eavesdropper Eve intercepts the particles sent among Alice and the
participants and then uses a unitary operation ${U_E}$ to entangle an ancillary
state $\left| E \right\rangle $ on the transmitted particle. In order to steal
secret information by measuring the ancillary state, Eve act the
unitary operator ${U_E}$ on $\left| E \right\rangle $ and the transmitted
particle. To simplify the description, we consider the bases corresponding to $j = 0$,
namely, $\left| {v_l^{\left( 0
\right)}} \right\rangle  = \frac{1}{{\sqrt d }}\sum\limits_{k = 0}^{d - 1}
{{w^{kl}}\left| k \right\rangle } $, so

\begin{equation}
{U_E}\left| k \right\rangle \left| E \right\rangle  = \sum\limits_{h = 0}^{d -
1} {{a_{kh}}\left| h \right\rangle \left| {{e_{kh}}} \right\rangle },
\label{eq:13}
\end{equation}

\begin{equation}
\begin{array}{l}
{U_E}\left| {v_l^{\left( 0 \right)}} \right\rangle \left| E \right\rangle  =
{U_E}\left( {\frac{1}{{\sqrt d }}\sum\limits_{k = 0}^{d - 1} {{w^{kl}}\left| k
\right\rangle } } \right)\left| E \right\rangle \\
\;\;\;\;\;\;\;\;\;\;\;\;\;\;\;\;\;\;\;\;\;= \frac{1}{{\sqrt d }}\sum\limits_{k = 0}^{d - 1}
{{w^{kl}}} \left( {\sum\limits_{h = 0}^{d - 1} {{a_{kh}}\left| h \right\rangle
\left| {{e_{kh}}} \right\rangle } } \right)\\
\;\;\;\;\;\;\;\;\;\;\;\;\;\; \;\;\;\;\;\;\;= \frac{1}{{\sqrt d }}\sum\limits_{k = 0}^{d - 1}
{\sum\limits_{h = 0}^{d - 1} {{w^{kl}}{a_{kh}}\left( {\frac{1}{{\sqrt d
}}\sum\limits_{m = 0}^{d - 1} {{w^{ - hm}}\left| {v_m^{\left( {\rm{0}} \right)}}
\right\rangle } } \right)} } \left| {{e_{kh}}} \right\rangle \\
\;\;\;\;\;\;\;\;\;\;\;\;\;\; \;\;\;\;\;\;\;= \frac{1}{d}\sum\limits_{k = 0}^{d - 1}
{\sum\limits_{h = 0}^{d - 1} {\sum\limits_{m = 0}^{d - 1} {{w^{kl -
hm}}{a_{kh}}\left| {v_m^{\left( {\rm{0}} \right)}} \right\rangle } } } \left|
{{e_{kh}}} \right\rangle,
\end{array}
\label{eq:14}
\end{equation}

{\noindent {\raggedright
where $w = {e^{\frac{{2\pi i}}{d}}}$, $\left| E \right\rangle $ is the initial
state of the auxiliary space, $\left| {{e_{kh}}} \right\rangle$ are pure ancillary states
determined uniquely by the unitary operation ${U_E}$, so
}}

\begin{equation}
\sum\limits_{h = 0}^{d - 1} {{{\left| {{a_{kh}}} \right|}^2} = 1,k \in \left\{
{0,1, \cdots ,d - 1} \right\}}.
\label{eq:15}
\end{equation}

{\noindent {\raggedright
For the sake of avoiding the rising error rate, Eve has to set $a{}_{kh} = 0$, $k,h
\in \left\{ {0,1, \cdots ,d - 1} \right\},k \ne h$. Therefore, (11) And (12) can be
simplified to
}}

\begin{equation}
{U_E}\left| k \right\rangle \left| E \right\rangle  = {a_{kk}}\left| k
\right\rangle \left| {{e_{kk}}} \right\rangle,
\label{eq:16}
\end{equation}

\begin{equation}
{U_E}\left| {v_l^{\left( 0 \right)}} \right\rangle \left| E \right\rangle  =
\frac{1}{d}\sum\limits_{k = 0}^{d - 1} {\sum\limits_{m = 0}^{d - 1} {{w^{k(l -
m)}}{a_{kk}}\left| {v_m^{\left( {\rm{0}} \right)}} \right\rangle } } \left|
{{e_{kk}}} \right\rangle.
\label{eq:17}
\end{equation}

{\noindent Similarly, to avoid the eavesdropping check, Eve has to set}

\begin{equation}
\sum\limits_{k = 0}^{d - 1} {{w^{k(l - m)}}{a_{kk}}\left| {{e_{kk}}}
\right\rangle }  = 0,
\label{eq:18}
\end{equation}

{\noindent where $m \in \left\{ {0,1, \cdots ,d - 1} \right\},m \ne l$. For any $l \in
\left\{ {0,1, \cdots ,d - 1} \right\}$, we can obtain $d$ equations}

\begin{equation}
{a_{00}}\left| {{e_{00}}} \right\rangle  = {a_{11}}\left| {{e_{11}}}
\right\rangle  =  \cdots  = {a_{d - 1,d - 1}}\left| {{e_{d - 1,d - 1}}}
\right\rangle.
\label{eq:19}
\end{equation}

{\noindent So, whatever quantum state Eve uses, he can only get the same information from
the auxiliary particles.{\scriptsize  } Similar analysis can be used for the other
quantum states $\left| {v_l^{\left( j \right)}} \right\rangle  = \frac{1}{{\sqrt d
}}\sum\limits_{k = 0}^{d - 1} {{w^{k\left( {l + jk} \right)}}\left| k
\right\rangle } $, so the entanglement measurement attack is
invalid in our scheme.
}

\subsubsection{Intercept and resend attack }

The eavesdropper Eve intercepts the transmitted particles among Alice and
the participants and resends some forged particles.{\scriptsize  } For a simple
description, we suppose that the eavesdropper Eve intercepts the quantum state
${\left| \phi  \right\rangle _k}$ sent by $Bob_k^{\left( i \right)}$ to $Bob_{k +
1}^{\left( i \right)}$. However, he does not know any information about the
measurement bases and only chooses the correct measurement bases with the
probability of $\frac{1}{d}$ to get measure outcome

\begin{equation}
{p_0} + \sum\limits_{i = 1}^k {{p_i}}.
\label{eq:20}
\end{equation}

{\noindent Even if the result is measured with the probability of $\frac{1}{d}$, the secret
information cannot be obtained because ${p_i},i \in \left\{ {k + 1, \cdots ,m}
\right\}$ is unknown. If Alice shares $n$ secret information, the
probability that eavesdropper succeed will be ${\left( {\frac{1}{d}} \right)^n}$.
With the increase of the number of $n$, there will be $\mathop {\lim }\limits_{n
\to \infty } {\left( {\frac{1}{d}} \right)^n} = 0$. The other is that Eve intercepts the $s_i$
sent by Alice to the participants, but the $s_i$ does not carry any information of the secret.
In short, Eve cannot obtain
the secret in intercept-and-resend attack.}

\subsubsection{Forgery attack}
If Alice shares a fake ${s_t}$ to $Bob_t^{\left( i \right)}$, the secret $s$
will not be restored by the participants in ${A_i}$. If one or some of the participants perform the false unitary
operation, they will be found by Alice because the measurement result will be inconsistent with Alice's expectation.
What's more, even if some dishonest participants performed the fake unitary transformation and Alice successfully measured
the expected result, there is no use for this attack. Because the  recovered secret $s'$ with ${H_2}' = h\left( {s'} \right)
\ne H_2 = h\left( s \right)$ guaranteed. So, the forgery attack is useless.

\subsubsection{Collusion attack }

If the participants in ${B_i},{B_i} \subseteq {A_i}$ collude to restore the
secret, they must obtain the $s_k^{(i)}$ and $\lambda _k^{(i)}$  of each participant in the
authorization set ${A_i}$ to recover $S_i$. When ${B_i} \subseteq {A_i}$, they can not get the other's
secret share information, so this attack is unsuccessful.

\section{Example}

Here, we explain our scheme more clearly by giving an example.

{\noindent{\raggedright\textbf{Example 3} According to the MSP and the access structure $\Gamma $ in the
example 2, assuming Alice wants to share secret $s = {\rm{3}} \in {{\cal Z}_5}$ among
the four participants $Bo{b_1},Bo{b_2},Bo{b_3},Bo{b_4}$, she prepares a random
vector $\vec \rho  = {\left( {4,1,0,2} \right)^T}$ firstly and then computes $\vec
s = M\vec \rho $ = ${\left( {{s_1},{s_2},{s_3},{s_4}} \right)^T}$ = ${\left(
{2,2,1,1} \right)^T}$\textbf{{\scriptsize . }} Next, she sends ${s_i}${\scriptsize
 } to {\scriptsize  }$Bo{b_i}$, $(i = 1,2,3,4)$ via a secure channel and publishes $H_1 =
h\left( 4 \right)$, $H_2 =
h\left( 3 \right)$. Without losing generality, we assume that the participants in
${A_1}$ want to restore the secret. The dealer Alice prepares a state $\left| \phi
\right\rangle  = \left| {\varphi _0^0} \right\rangle  = \frac{1}{{\sqrt 5
}}\sum\limits_{i = 0}^4 {\left| i \right\rangle } $ and performs ${U_{{p_0},{q_0}}}
= {U_{3,2}}$ on it to obtain ${\left| \phi  \right\rangle _0} = {U_{3,2}}\left|
{\varphi _0^0} \right\rangle  = \left| {\varphi _3^2} \right\rangle $,
where {\scriptsize  }${p_0} = s = 3$ is the secret, ${q_0} = 2 \in {{\cal Z}_5}$ is a
randomly selected secret value only known by Alice. Next she sends the quantum
state ${\left| \phi  \right\rangle _0} = \left| {\varphi _3^2} \right\rangle $
to $Bo{b_1}$. After receiving ${\left| \phi  \right\rangle _0} = \left| {\varphi
_3^2} \right\rangle $, $Bo{b_1}$ performs the unitary operation ${U_{{\lambda
_1}{s_1},{\lambda_1}}} = {U_{2,1}}$ to get ${\left| \phi  \right\rangle _1} =
\left| {\varphi _0^3} \right\rangle $ and sends it to $Bo{b_2}$. When receiving
${\left| \phi  \right\rangle _1} = \left| {\varphi _0^3} \right\rangle $,
$Bo{b_2}$ performs the unitary operation${U_{{\lambda _2}{s_2},{\lambda _2}}} =
{U_{2,1}}$ to get ${\left| \phi  \right\rangle _2} = \left| {\varphi _2^4}
\right\rangle $ and sends to $Bo{b_3}$. After receiving ${\left| \phi
\right\rangle _2} = \left| {\varphi _2^4} \right\rangle $, $Bo{b_3}${\scriptsize
} performs ${U_{{\lambda _3}{s_3},{\lambda _3}}} = {U_{0,0}}$ to get ${\left| \phi
\right\rangle _3} = \left| {\varphi _2^4} \right\rangle $ and sends it to Alice.
For the authorization set ${A_1}$, when all participants act and transmit
particle, the final quantum state is

\begin{equation}
\left| {{\varphi }} \right\rangle_{final}  = \left( {\prod\limits_{i = 0}^3
{{U_{{p_i},{q_i}}}} } \right)\left| {\varphi _0^0} \right\rangle  = \left|
{\varphi _{\sum\limits_{i = 0}^3 {{p_i}} }^{\sum\limits_{i = 0}^3 {{q_i}} }}
\right\rangle  = \left| {\varphi _2^4} \right\rangle.
\label{eq:21}
\end{equation}

{\noindent In this case, Alice selects ${M_4} = \left\{ {\left| {\varphi _j^{(4)}}
\right\rangle \left| {j \in \left\{ {0,1,2,3,4} \right\}} \right.} \right\}$ to
measure ${\left| {{\varphi}} \right\rangle { _{final}}}$ $ = \left| {\varphi _2^4}
\right\rangle $ and records the measurement result ${r_1}$. Afterwards, Alice
checks whether ${r_1} = 2$ and $H_1=h(4)$ are true. If not, the scheme is terminated. If they are
established, the measurement result ${r_1}$ is sent to each participant in
${A_1}$ through a quantum secure channel. After the participant receives it, the secret $s$
can be recovered as}

\begin{equation}
s = p_0^{(1)} = {r_1} - p_1^{(1)} - p_2^{(1)} - p_3^{(1)} = {r_1} - \lambda
_1^{(1)}s_{\rm{1}}^{(1)} - \lambda _2^{(1)}s_2^{(1)} - \lambda _3^{(1)}s_3^{(1)}.
\label{eq:22}
\end{equation}

{\noindent That is $s = 2 - (2 + 2 + 0) = 3$. Last but not least, they can check $H_1$ to make certain of the  authenticity of the secret.}
}}

\section{Comparison}

In this section, we give a comparison among our scheme and other similar $d$-dimensional QSS schemes[24,35,36] in terms of basic properties, computational complexity and communication costs. The schemes in Ref.[24, 36] are the threshold QSS, however the scheme in Ref. [35] and ours are the general access structure QSS. The general access structure makes the level and influence of the participants different, making the scheme more flexible. They all use the Hash function to make the verifiability of the d-dimensional QSS scheme. The scheme proposed by Song et al.[24] shared a classical secret by utilizing polynomials according to the Lagrange interpolation formula. The transformation of the particles includes some operations such as d-level CNOT, QTF, Inverse QTF, and generalized Pauli operator. However, the general access structure QSS is far more flexible and practical than the threshold one. In Ref.[35], Mashhadi proposed a hybrid secret sharing based on the quantum Fourier transform and monotone span program, in which the participants recover the secret by means of measuring the entangled state. The number of unitary operators is not much different in the premise, while the number of required quantum states and the number of measurement operations are greatly reduced, which consumes less quantum resources and the scheme is more practical. Qin et al.[36] put forward a verifiable $(t,n)$ threshold QSS using $d$-dimensional Bell state and they realize the authentication of quantum state transmission by adding some decoy particles. According to the Lagrange interpolation and the unitary operation, they can recover the secret with measuring the final Bell state. The Specific comparison of basic property among Ref.[24,35,36] and ours is given in Table 1. The comparison of the computational complexity and communication costs of the general access structure QSS[35] and the new is given in Table 2.

{\raggedright
\begin{center}

\begin{table}[h]
\caption{Basic comparison among the QSS schemes}

\vspace{3pt} \noindent
\begin{tabular}{p{54pt}p{82pt}p{65pt}p{62pt}p{58pt}}
\hline
\parbox{54pt}{\raggedright
{Property}
} & \parbox{82pt}{\raggedright
{Song[24]}
} & \parbox{65pt}{\raggedright
{Mashhadi[35]}
} & \parbox{62pt}{\raggedright
{Qin[36]}
} & \parbox{76pt}{\raggedright
{New}
} \\
\hline
\parbox{54pt}{\raggedright
{Model}
} & \parbox{82pt}{\raggedright
{$(t,n)$threshold}
} & \parbox{65pt}{\raggedright
{General}
} & \parbox{62pt}{\raggedright
{$(t,n)$threshold}
} & \parbox{76pt}{\raggedright
{General}
} \\
\parbox{54pt}{\raggedright
{Verification}
} & \parbox{82pt}{\raggedright
{Hash function}
} & \parbox{65pt}{\raggedright
{Hash function}
} & \parbox{62pt}{\raggedright
{Hash function}
} & \parbox{76pt}{\raggedright
{Hash function}
} \\
\parbox{54pt}{\raggedright
{Secret}
} & \parbox{82pt}{\raggedright
{Classic}
} & \parbox{65pt}{\raggedright
{Classic}
} & \parbox{62pt}{\raggedright
{Classic}
} & \parbox{76pt}{\raggedright
{Classic}
} \\
\parbox{54pt}{\raggedright
{Dimension}
} & \parbox{82pt}{\raggedright
{$d$}
} & \parbox{65pt}{\raggedright
{$d$}
} & \parbox{62pt}{\raggedright
{$d$}
} & \parbox{76pt}{\raggedright
{$d$}
} \\
\parbox{54pt}{\raggedright
{Method}
} & \parbox{82pt}{\raggedright
{LI}
} & \parbox{65pt}{\raggedright
{MSP,LC}
} & \parbox{62pt}{\raggedright
{LI}
} & \parbox{76pt}{\raggedright
{MSP,MUB,LC}
} \\
\parbox{54pt}{\raggedright
{NQO}
} & \parbox{82pt}{\raggedright
$QFT${,$QF{T^{ - 1}}$,\\Pauli}
} & \parbox{65pt}{\raggedright
$QFT${,Pauli}
} & \parbox{62pt}{\raggedright
{UO}
} & \parbox{76pt}{\raggedright
{UT}
} \\
\hline
\end{tabular}
\vspace{2pt}
\end{table}

\end{center}

\begin{center}

\begin{table}[h]
{\noindent\caption{Comparison of communication costs and computational complexity}}

\vspace{3pt} \noindent
\begin{tabular}{p{160pt}p{90pt}p{70pt}}
\hline
\parbox{128pt}{\raggedright
{Property}
} & \parbox{128pt}{\raggedright
{Mashhadi[35]}
} & \parbox{128pt}{\raggedright
{Ours}
} \\
\hline
\parbox{128pt}{\raggedright
{Number of message particles}
} & \parbox{128pt}{\raggedright
{$m-1$}
} & \parbox{128pt}{\raggedright
{$1$}
} \\
\parbox{128pt}{\raggedright
{Unitary operation}
} & \parbox{128pt}{\raggedright
{$m$}
} & \parbox{128pt}{\raggedright
{$m+1$}
} \\
\parbox{128pt}{\raggedright
{$QTF$}
} & \parbox{128pt}{\raggedright
{1}
} & \parbox{128pt}{\raggedright
{$-$}
} \\
\parbox{128pt}{\raggedright
{Measure operation}
} & \parbox{128pt}{\raggedright
{$m$}
} & \parbox{128pt}{\raggedright
{$1$}
} \\
\parbox{128pt}{\raggedright
{Hash function}
} & \parbox{128pt}{\raggedright
{$2$}
} & \parbox{128pt}{\raggedright
{$2$}
} \\
\hline
\end{tabular}
\vspace{2pt}
\end{table}

\end{center}
\begin{remark}
{\noindent LI: Lagrange interpolation, MSP: Monotone span program, MUB: Mutually unbiased bases, LC: Linear computation, NQO: necessary quantum operation, QTF: Quantum Fourier Transform, ${QFT^{-1}}$: Inverse
Quantum Fourier Transform, UO: Unitary operation, UT: Unitary transformation.}
\end{remark}

\section{Conclusions}

The verifiable quantum secret sharing scheme based on the access structure is
very useful in practice. In this paper, we construct a verifiable quantum secret
sharing scheme based on the property of the mutually unbiased base and the
monotone span program. The dealer and participants in the authorization set can
restore secret through the transformation and transmission of a $d$-dimensional
quantum state as well as linear secret sharing. In addition, the correctness, verifiability and security analysis
of the scheme have been proved. Finally, a specific example and a comparison are given to further clarify the advantages and practicality of our scheme.

For the future work,the verifiability of the scheme is analyzed from the view that
the recovered secret is consistent with the original one. However, the issue of
mutual authentication among the participants in the authorization set is still
worth studying.

%
%
%
%

\end{document}